\documentclass[a4paper]{article}
\usepackage{float}
\usepackage{INTERSPEECH2022}
\usepackage{stfloats}
\usepackage{hyperref}
\usepackage{balance}

\title{MOSRA: Joint Mean Opinion Score and Room Acoustics \\ Speech Quality Assessment}
\name{Karl El Hajal$^{1,2}$, Milos Cernak$^1$, Pablo Mainar$^1$}
\address{
  $^1$Logitech Europe S.A., Lausanne, Switzerland \\
  $^2$\'Ecole Polytechnique F\'ed\'erale de Lausanne (EPFL), Lausanne, Switzerland}
\email{karl.elhajal@epfl.ch, mcernak@logitech.com, pmainar@logitech.com}

\begin{document}

\maketitle
\begin{abstract}

The acoustic environment can degrade speech quality during communication (e.g., video call, remote presentation, outside voice recording), and its impact is often unknown. Objective metrics for speech quality have proven challenging to develop given the multi-dimensionality of factors that affect speech quality and the difficulty of collecting labeled data. Hypothesizing the impact of acoustics on speech quality, this paper presents MOSRA: a non-intrusive multi-dimensional speech quality metric that can predict room acoustics parameters (SNR, STI, T60, DRR, and C50) alongside the overall mean opinion score (MOS) for speech quality. By explicitly optimizing the model to learn these room acoustics parameters, we can extract more informative features and improve the generalization for the MOS task when the training data is limited. Furthermore, we also show that this joint training method enhances the blind estimation of room acoustics, improving the performance of current state-of-the-art models. An additional side-effect of this joint prediction is the improvement in the explainability of the predictions, which is a valuable feature for many applications.

\end{abstract}


\section{Introduction}

Speech quality is becoming increasingly important in daily life with the rise of video-conferencing solutions, digital content creation, collaborative remote gaming, and other audio applications. However, it has been traditionally challenging to measure it, especially when no clean reference is available. One reason is that many audio dimensions can influence speech quality. For example, it can be affected by reverberant rooms \cite{zhao2018two}, codecs \cite{zango_codec}, network losses \cite{lingfen_packetloss}, or additive noises \cite{arehart_noise}, among others.

The mean opinion score (MOS) belongs to the primary evaluation metrics of interest. The main method to assess MOS on speech quality is subjective listening tests, as described by the ITU-T P.800~\cite{p800}. The obvious drawback is that it is time-consuming and expensive to carry out subjective tests. More recently, crowd-sourcing has been used to evaluate speech samples \cite{p808}, but still, it is impractical and costly. Hence the search for objective metrics for speech quality. The main industry standards used in the last years include PESQ \cite{pesq}, POLQA \cite{polqa} or 3-QUEST \cite{3quest}. All of them have been validated for specific use cases and might not cover the entire range of distortions present in today’s applications \cite{nessler21_interspeech}. In addition, they are intrusive methods, meaning that they require a clean reference signal to make quality predictions. 

Development of non-intrusive metrics for speech quality has been a hot research topic in recent years, especially with the rise of deep learning. Indeed, there is now a large variety of models (CNNs \cite{nessler21_interspeech, wawenets} , BLSTM \cite{quality_net}, CNN-LSTM \cite{nisqa-v1}, CNN-Self Attention \cite{nisqa-v2}, CNN-BLSTM-Self Attention \cite{dong_mos_multitask_attention}) and input features for the models (raw audio \cite{wawenets}, log-mel-spectrograms \cite{nisqa-v1, nisqa-v2}, MFCCs \cite{nessler21_interspeech, microsoft_nonintrusive}).

Training these models generally requires a considerable effort in data collection. Many researchers have tried to use joint-learning to tackle this problem of lower data availability. The strategy is to learn speech quality labels jointly with other audio dimensions, hoping that the model will learn to extract more informative features from the audio without requiring as much data. In NISQA  \cite{nisqa-v2}, the model was trained to predict four perceptual speech quality dimensions (noisiness, coloration, discontinuity, loudness), in addition to the speech quality itself. An additional advantage of this method is the improved explainability of the model. MetricNet \cite{metricnet} explores adding a signal reconstruction branch that aims to reconstruct the clean signal from the degraded signal for which we are predicting the MOS score. Another multi-task model learns to jointly predict four objective speech quality and intelligibility scores (PESQ, ESTOI, HASQI, SDR)~\cite{dong_mos_multitask_attention}. Finally, joint speech intelligibility prediction and scene classification is presented in~\cite{multitask-intelligibility-scene}. 

Room acoustics is potentially one of the main factors affecting speech quality. A different line of research thus has proven possible to predict room acoustics parameters blindly from audio samples. For example, \cite{room_acoustics_1} trains a Convolutional Recurrent Neural Network to jointly and blindly estimate reverberation time (T60), clarity (C50 and C80), and direct-to-reverberant ratio (DRR) for both music and speech. In \cite{room_acoustics_2} the authors expand the model to also predict the speech transmission index (STI) and signal-to-noise ratio (SNR).

Therefore, we hypothesize that training a model which is explicitly optimized to capture these multi-dimensional factors will improve the performance of the quality predictions, especially when a small amount of quality data is available. Equivalently, we also hypothesize that adding quality MOS will also improve the prediction of room acoustics parameters. Finally, having a joint model improves the explainability of the predictions, which is an essential feature for many applications.

This work combines speech quality and room acoustics data and trains a model to predict them jointly. We call this model MOSRA (MOS + Room Acoustics). We evaluate its performance against current state-of-the-art methods on non-intrusive speech quality metrics and blind room acoustics predictors. The paper is structured as follows. The next section introduces our model in detail. Section~\ref{sec:evaluation} describes the evaluation datasets that we use and the baselines against which we compare our model. Finally, sections~\ref{sec:results} and \ref{sec:discussion} show the results and their discussion, respectively.

\section{Methods}
\label{sec:methods}

\begin{figure}
    \centering
    \includegraphics[width = .9\columnwidth]{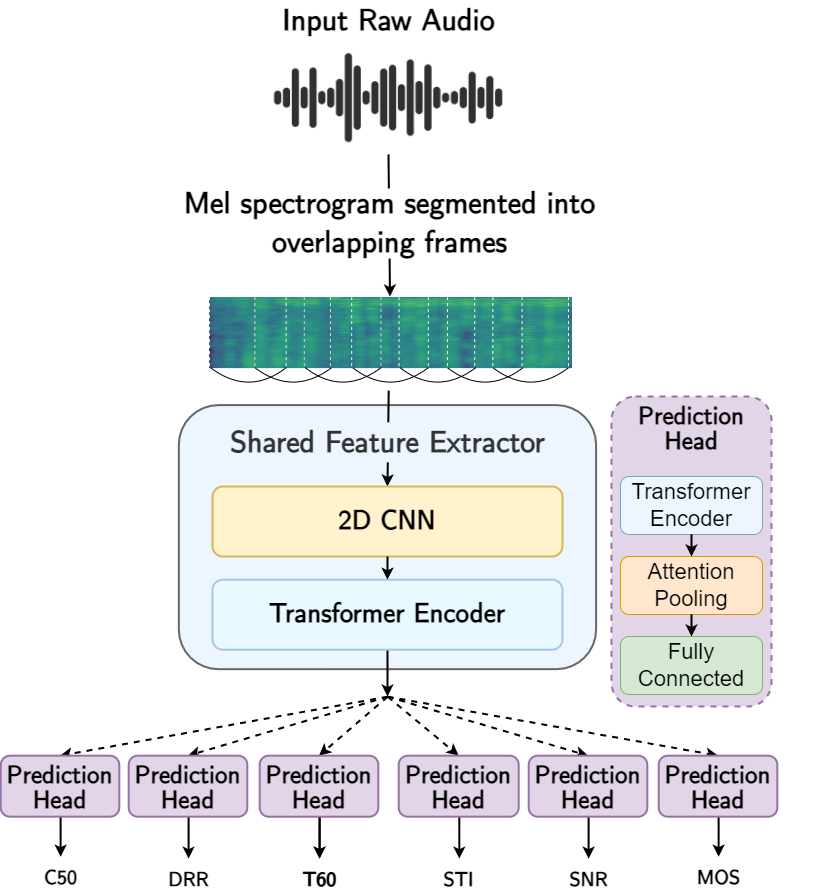}
    \caption{Overview of the MOSRA model}
    \label{fig:model_architecture}
\end{figure}

\begin{figure}
    \centering
    \includegraphics[width = \columnwidth]{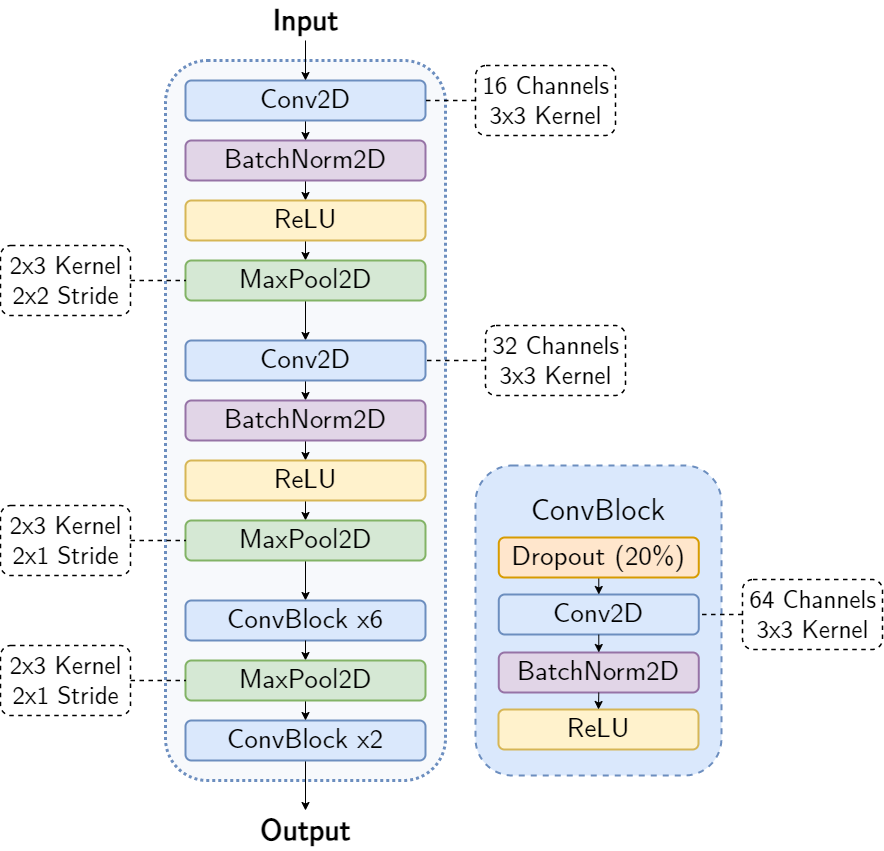}
    \caption{Overview of the CNN that extracts features from individual segments}
    \label{fig:CNN_architecture}
\end{figure}

This section introduces our proposed MOSRA model, the architecture of which is shown in Fig. \ref{fig:model_architecture}.
It is designed to be able to learn and predict several tasks jointly, which, as we will later discuss, can improve the features that are extracted in the shared parts of the model.
The different stages of the model are described in detail below.

\subsection{Shared feature extractor}

The input features used are Mel-spectrogram segments inspired by \cite{nisqa-v2}. These are obtained by first computing the Mel-spectrogram of the input audio file with 48 Mel bands, an FFT window size of 20ms, and a hop size of 10ms. The maximum Mel band frequency is set to 20kHz, which means that full band audio is supported. The spectrogram is then separated into overlapping frames, each with a width of 150ms and a hop size of 40ms.

A CNN, described in Fig. \ref{fig:CNN_architecture}, is then used to extract features from individual segments. Its architecture is inspired by \cite{nisqa-v1} but has additional blocks to increase its capacity.

The features extracted from each frame are then passed as input to a Transformer encoder \cite{vaswani} to extract temporal dependencies between the segments across time. The encoder has $M = 2$ layers, $h = 1$ head, intermediate fully-connected layers of dimension $d_{ff} = 64$, and accepts inputs with $d_{\text{model}} = 64 $ features.

\subsection{Prediction heads}


The outputs from the shared feature extractor are then fed to individual heads, each of which is entrusted with predicting for a single task. 

Each head first has its own Transformer encoder to further extract temporal dependencies that are of interest for each specific task. Each encoder has $M = 1$ layer, $h = 1$ head, $d_{ff} = 32$, and $d_{\text{model}} = 32 $.

The encoder's outputs are then passed to an attention pooling block, as described in \cite{nisqa-v2}, and then finally to a fully-connected layer which outputs the final prediction.

\subsection{Tasks}

We jointly train the model to predict MOS, SNR, and the following room acoustics estimators: T60, STI, DRR, and C50. Room acoustics are some of the main factors that can affect speech quality, and there isn't sufficient data with MOS labels that covers the effects of room acoustics on quality. This gives us the incentive to try to learn these tasks jointly to improve the model's ability to generalize to speech samples that are fairly affected by acoustics. 

\subsection{Loss function}

For each task, we first calculate its MSE loss:

\begin{equation}
    \text{MSE} = \sum_{i=1}^{n}(y_{pred}-y_{true})^2
\end{equation}

The overall loss is then a weighted sum of the per-task losses:

\begin{equation}
    \text{Loss}_\text{room acoustics} = \text{MSE}_\text{SNR} + \text{MSE}_\text{T60} + \text{MSE}_\text{STI} + \text{MSE}_\text{DRR} + \text{MSE}_\text{C50}
\end{equation}

\begin{equation}
    \text{Loss}_\text{overall} = \lambda_{1} \text{MSE}_\text{MOS} + \lambda_{2} \text{Loss}_\text{room acoustics}
\end{equation}

Since the room acoustics estimators have very different scales, we normalize their labels and network outputs by calculating the mean and standard deviation of each estimator across the dataset. We note that, in this study, we did not explore tuning individual weights for each estimator, but it would be worth exploring in the future, along with adding additional estimators and trying to find the best selection of dimensions.

The task weights $\lambda_{1}$ and $\lambda_{2}$ are hyper-parameters that are used, in a sense, to give different levels of priority for the different tasks \cite{caruana_multitask_learning}. Here, we give priority to the MOS task, giving it a weight $\lambda_{1} = 2$, while scaling down the rest of the tasks with a weight of $\lambda_{2} = \frac{1}{5}$. The intuition behind the $\lambda_{2}$ value was to give the five room acoustics tasks as much weight as the single MOS task since the latter is the main one we are interested in. $\lambda_{1}$ was set to $2$, following some experiments that showed that increasing the weight of the MOS task improved its performance without affecting the other tasks.

\subsection{Training}

The model was trained using the ADAM \cite{kingma2014adam} optimizer, a learning rate  $\alpha = 5\mathrm{e}{-4}$, a batch size of 32, and early stopping with patience set to 15. 

We use separate datasets for the MOS and room acoustics characterization tasks in this work. Therefore, given that we would like the model to learn all the tasks jointly, we opt to interleave MOS data, and room acoustics data \cite{caruana_multitask_learning}. This means that the model performs a forward pass on a batch of MOS data every iteration and then performs another forward pass on a batch of room acoustics data. The MOS results from the first pass and the room acoustics results from the second one are then used to calculate the loss for that iteration.

Finally, the MSE loss for the MOS task is the metric used for early stopping. Indeed, when training on multiple tasks, the tasks can converge at different speeds \cite{caruana_multitask_learning}, which makes it unfeasible most of the time to find an optimal stopping point for all tasks. Therefore, we give priority to the MOS task in this case and stop training when its validation performance stops improving.

\section{Experimental setup}
\label{sec:evaluation}

We evaluate models that are trained on two different sets of data. First, we discuss models trained on what we will refer to as the \textit{Smaller Data}, which has a relatively small amount of speech quality labels. This constraint gave us the incentive to experiment with joint learning and to assess whether it will lead to models that generalize better. We then further evaluate models that are trained on the \textit{Challenge Data}, which was released for the ConferencingSpeech 2022\footnote{\href{https://tea-lab.qq.com/conferencingspeech-2022}{https://tea-lab.qq.com/conferencingspeech-2022}}  challenge \cite{challenge_evaluation_plan} and which significantly increases the amount of speech quality labels available. We consequently assess the effect that training the model with the extra data has on performance. 
We describe the different datasets and baselines below. 

\subsection{Datasets}

\subsubsection{Smaller Data}
The smaller set of data consists of the following:

For the MOS task, the NISQA \cite{nisqa-v2} training datasets, consisting of 11,020 speech files that are between 6 and 12 seconds long, are used for training. The NISQA validation sets are further used for validation when training the model. We finally test our models on the NISQA test datasets, and on the internal ReverbSpeechQuality dataset which was described in \cite{nessler21_interspeech}. The latter contains a wide range of reverberations that are applied to the speech. This focus on room acoustics-related effects on quality distinguishes it from other datasets which focus more on degradations that are caused by issues with the communication channel.

For the room acoustics characterization tasks, we prepare and use a dataset as described in \cite{room_acoustics_1, room_acoustics_2}. The training set consists of 80k samples with SNR, STI, RT60, C50, C80, and DRR labels, while the validation and test sets consist of 10k samples each.

\subsubsection{Challenge Data}
For the challenge, two additional training datasets were open sourced. The Tencent Corpus contains 10k speech samples without reverberation, and 4k samples that include reverberation. 80\% of it is used for training, and the rest is used as a test set for the challenge.  The PSTN \cite{pstn} training dataset consists of around 62k samples, 95\% of which is used for training while the rest is also used as a challenge test set. Additionally, the TUB evaluation set, consisting of 433 samples, is used to assess models for the challenge.

Our model is optimized to work with audio files that have a sample rate of 48kHz. Therefore, any data that is used which has a smaller sampling rate is upsampled before being fed to the model.

\subsection{Baseline systems}
We compare our model to several baselines. We first discuss the models trained on the smaller set of data and then move on to others that are trained on the challenge data.

\subsubsection{Smaller Data}
First, for the MOS estimation task, we compare the performance difference between the multi-dimensional MOSRA model, and the same architecture trained to predict MOS only. We then compare the multi-dimensional model with the open-source multi-dimensional NISQA model from \cite{nisqa-v2}.  

For the room acoustics characterization tasks, we compare our model with the state of the art ``universal room acoustics estimator'' (URAE) model presented in \cite{room_acoustics_2}.

\subsubsection{Challenge Data}
The baseline used for the challenge consists of a variation of the NISQA model trained on the challenge data for 50 epochs. We thus compare it with MOSRA models trained on the challenge data. We further compare them with POLQA \cite{polqa}, which is the current ITU-T recommendation and is intrusive.

\subsection{Performance metrics}
To evaluate the MOS performance of each model, we calculate the Pearson Correlation Coefficient and RMSE after mapping with a third-order polynomial function, which are included in the ITU-T P.1401 recommendation \cite{p1401}. They are computed for each dataset independently due to the subjective nature of the tests. 

To evaluate the performance on the room acoustics characterization tasks, we use the RMSE.

\begin{table}[b]
\centering
\caption{PCC ($\uparrow$) and RMSE after third-order mapping ($\downarrow$) comparison of MOSRA models trained on the Smaller Data}
\label{tab:mos_results_original_data1}
\resizebox{\columnwidth}{!}{
\begin{tabular}{@{}lllrlrlrlrlrlrlr@{}}
\toprule
Dataset   & \# Files &  \multicolumn{2}{c}{\begin{tabular}[c]{@{}c@{}}MOSRA \\ MOS only \end{tabular}} & \multicolumn{2}{c}{\begin{tabular}[c]{@{}c@{}}MOSRA \\ Multi-dim\end{tabular}}  \\

\cmidrule(rl){3-4} \cmidrule(rl){5-6} 
  &   & PCC & RMSE &  PCC  & RMSE \\
  \midrule
NISQA\_TEST\_LIVETALK &   232&     0.63   &  0.73&  \textbf{ 0.68} & \textbf{0.69}  \\
NISQA\_TEST\_FOR  &   240&   0.85   & 0.45 &\textbf{0.87}& \textbf{0.42}   \\
NISQA\_TEST\_P501 &   240&   \textbf{0.9}   & \textbf{0.44} &0.89& 0.47  \\
ReverbSpeechQuality &   4032&  0.77 & 0.59  & \textbf{0.85}   & \textbf{0.48 } \\ \bottomrule
\end{tabular}
}
\end{table}

\section{Results}
\label{sec:results}

\subsection{Speech Quality}

\subsubsection{Smaller Data}

Table \ref{tab:mos_results_original_data1} first shows the comparison between the multi-dimensional MOSRA model and a MOS only variation, both trained on the smaller set of data. We can observe that the multi-dimensional model outperforms the single-task model in all but one of the cases, most significantly in the \textit{ReverbSpeechQuality} test.

\begin{table}
\centering
\caption{PCC ($\uparrow$) and RMSE after third-order mapping ($\downarrow$) comparison of MOSRA and NISQA trained on the Smaller Data}
\label{tab:mos_results_original_data2}
\resizebox{\columnwidth}{!}{
\begin{tabular}{@{}lllrlrlrlrlrlrlr@{}}
\toprule
Dataset   & \# Files &  \multicolumn{2}{c}{\begin{tabular}[c]{@{}c@{}}NISQA \\ Multi-dim\end{tabular}} &  \multicolumn{2}{c}{\begin{tabular}[c]{@{}c@{}} MOSRA \\ Multi-dim\end{tabular}}  \\

\cmidrule(rl){3-4} \cmidrule(rl){5-6} 
  &   & PCC & RMSE &  PCC  & RMSE   \\
  \midrule
NISQA\_TEST\_LIVETALK &   232&  \textbf{0.74} & \textbf{0.63}   &     0.68 & 0.69  \\
NISQA\_TEST\_FOR  &   240&  \textbf{0.88} & \textbf{0.41} &  0.87& 0.42   \\
NISQA\_TEST\_P501 &   240&   \textbf{0.92} & \textbf{0.4}  &  0.89& 0.47  \\
ReverbSpeechQuality &   4032&   0.77  &   0.58 &   \textbf{0.85}   & \textbf{0.48 } \\ \bottomrule
\end{tabular}
}
\end{table}

\begin{table}
\centering
\caption{RMSE ($\downarrow$) performance comparison on the room acoustics characterization tasks}
\label{tab:room_acoustics_results}
\resizebox{\columnwidth}{!}{
\begin{tabular}{@{}llllll@{}}
\toprule
  Model & \begin{tabular}[c]{@{}l@{}}SNR\\ {[}dB{]}\end{tabular} & STI   & \begin{tabular}[c]{@{}l@{}}DRR\\ {[}dB{]}\end{tabular} & \begin{tabular}[c]{@{}l@{}}T60\\ {[}s{]}\end{tabular} & \begin{tabular}[c]{@{}l@{}}C50\\ {[}dB{]}\end{tabular} \\ \midrule
URAE & 3.36   & 0.08  & 4.6& 0.46  & 9.60   \\
MOSRA Room acoustics & \textbf{2.63} & 0.073  & 4.21 &  0.45 & 8.84 \\
MOSRA Multi-dim & 3.22   & \textbf{0.072} & \textbf{4.15}   & \textbf{0.44}  & \textbf{8.62}   \\ 
\bottomrule
\end{tabular}
}
\end{table}

We also compare the multi-dimensional MOSRA with the multi-dimensional NISQA model in Table \ref{tab:mos_results_original_data2}. We observe that NISQA performs better than MOSRA on the NISQA test datasets, while MOSRA performs better on the ReverbSpeechQuality dataset.

\subsubsection{Challenge Data}

\begin{table*}[bp]
\centering
\caption{PCC ($\uparrow$) and RMSE after third-order mapping ($\downarrow$) comparison of models trained on the Challenge data}
\label{tab:mos_results_challenge_data}
\begin{tabular}{@{}lllrlrlrlrlrlrlr@{}}
\toprule
Dataset   & \# Files & \multicolumn{2}{c}{POLQA} & \multicolumn{2}{c}{\begin{tabular}[c]{@{}c@{}} Challenge \\ Baseline\end{tabular}} & \multicolumn{2}{c}{\begin{tabular}[c]{@{}c@{}} MOSRA \\ MOS only\end{tabular}} & \multicolumn{2}{c}{\begin{tabular}[c]{@{}c@{}} MOSRA \\ Multi-dim\end{tabular}} \\

\cmidrule(rl){3-4} \cmidrule(rl){5-6} \cmidrule(rl){7-8} \cmidrule(rl){9-10} 
  &   & PCC & RMSE &  PCC  & RMSE  & PCC & RMSE & PCC & RMSE \\
  \midrule
NISQA\_TEST\_LIVETALK &   232&  n/a  &  n/a  & 0.43 & 0.85 & \textbf{0.76} & \textbf{0.61}  &   0.74  & 0.63  \\
NISQA\_TEST\_FOR  &   240& 0.85 &   0.45   & 0.70 & 0.62 &  0.85 &   0.46 &   \textbf{0.86} & \textbf{0.44} \\
NISQA\_TEST\_P501 &   240& 0.88   & 0.49  &  0.63 & 0.78  & 0.86 &   0.48   & \textbf{0.9} & \textbf{0.45}  \\
ReverbSpeechQuality &   4032&  0.78 & 0.53   & 0.82  & 0.51 & \textbf{0.89} & \textbf{0.41} & 0.88  & 0.43 \\ 
\midrule
Tencent Eval & 2898 &  &  & 0.881 & 0.55 &  &  & \textbf{0.941} & \textbf{0.393} \\
TUB Eval & 433 &  &  & 0.348 & 0.649 &  &  & \textbf{0.535} & \textbf{0.589} \\
PSTN Eval & 1039 &  &  & 0.361 & 0.293 &  & & \textbf{0.534} & \textbf{0.271} \\
\bottomrule
\end{tabular}
\end{table*}

Table \ref{tab:mos_results_challenge_data} first shows the results for the MOSRA models that are trained on the challenge data. In this case, we can no longer observe a clear difference between the two models, as they perform similarly on all test sets.

We also compare MOSRA with the challenge baseline. The baseline is clearly outperformed by the MOSRA models. We note that for the challenge evaluation sets, we can only present the results for the baseline and our Multi-dim submission.

Finally, POLQA is also outperformed by the MOSRA models. The \textit{NISQA\_TEST\_LIVETALK} dataset does not have clean reference files alongside the degraded ones, meaning that POLQA cannot perform predictions for this set.

\subsection{Room Acoustics Characterization}

Table \ref{tab:room_acoustics_results} shows the results for the room acoustics estimation tasks. MOSRA consistently outperforms the URAE model on the five tasks. Further, the model trained to only predict the room acoustics tasks is better at predicting SNR, while it is outperformed by the multi-dimensional model on the four other tasks.

\section{Discussion \& conclusions}
\label{sec:discussion}

Training our model jointly on MOS and room acoustics characterization tasks significantly helped the model's performance when the subjective quality data was scarce, notably when predicting for samples that are greatly affected by room acoustics-related degradations such as heavy reverberation. 

However, that advantage was not as clear when the models were trained on the much larger amounts of speech quality data that were released for the challenge. This might either be because the extra data allowed the model to extract the features learned from the room acoustics characterization tasks, or because the model was not designed with enough capacity to handle the substantial data increase. Indeed, the model's size (410k parameters) is relatively small and was designed with the smaller set of data in mind. 
Nonetheless, we still observe a clear improvement over the baseline on the challenge evaluation datasets. At the same time, the model's small size and architecture make it time and space efficient, and therefore easy to deploy in many practical settings. On average, it needs 120ms and 30MB to process an 8s long clip on a single threaded Intel Core i7 clocked at 3.6 GHz.

Further, the model's performance on the room acoustics characterization tasks successfully improves on the state of the art, showing the versatility of this architecture. The results also confirmed our hypothesis that jointly training the model to predict MOS and the room acoustics tasks would improve its performance all-round. Indeed, the model's shared feature extractor seems to benefit from all the extra data that it's trained on thanks to joint learning, and is consequently able to produce features that contain more useful information.

Finally, having the additional tasks adds a layer of explainability to the model's predictions, which is an important feature in many use cases.


\clearpage

\bibliographystyle{IEEEtran}
\balance
\bibliography{paperbib}

\begin{thebibliography}{10}
\providecommand{\url}[1]{#1}
\csname url@samestyle\endcsname
\providecommand{\newblock}{\relax}
\providecommand{\bibinfo}[2]{#2}
\providecommand{\BIBentrySTDinterwordspacing}{\spaceskip=0pt\relax}
\providecommand{\BIBentryALTinterwordstretchfactor}{4}
\providecommand{\BIBentryALTinterwordspacing}{\spaceskip=\fontdimen2\font plus
\BIBentryALTinterwordstretchfactor\fontdimen3\font minus
  \fontdimen4\font\relax}
\providecommand{\BIBforeignlanguage}[2]{{%
\expandafter\ifx\csname l@#1\endcsname\relax
\typeout{** WARNING: IEEEtran.bst: No hyphenation pattern has been}%
\typeout{** loaded for the language `#1'. Using the pattern for}%
\typeout{** the default language instead.}%
\else
\language=\csname l@#1\endcsname
\fi
#2}}
\providecommand{\BIBdecl}{\relax}
\BIBdecl

\bibitem{zhao2018two}
Y.~Zhao, Z.-Q. Wang, and D.~Wang, ``Two-stage deep learning for
  noisy-reverberant speech enhancement,'' \emph{IEEE/ACM transactions on audio,
  speech, and language processing}, vol.~27, no.~1, pp. 53--62, 2018.

\bibitem{zango_codec}
Y.~Zango, R.~Le~Bouquin~Jeannès, and C.~Quinquis, ``Modeling speech and audio
  codecs reverberation artifact,'' in \emph{2012 Proceedings of the 20th
  European Signal Processing Conference (EUSIPCO)}, 2012, pp. 2070--2074.

\bibitem{lingfen_packetloss}
L.~Sun and E.~Ifeachor, ``Perceived speech quality prediction for voice over
  ip-based networks,'' in \emph{2002 IEEE International Conference on
  Communications. Conference Proceedings. ICC 2002 (Cat. No.02CH37333)},
  vol.~4, 2002, pp. 2573--2577 vol.4.

\bibitem{arehart_noise}
K.~Arehart, J.~Kates, and M.~Anderson, ``Effects of noise, nonlinear
  processing, and linear filtering on perceived speech quality,'' \emph{Ear and
  Hearing}, vol.~31, no.~3, pp. 420--436, 2010.

\bibitem{p800}
{ITU-T Recommendation P.800}, ``Methods for objective and subjective assessment
  of quality,'' 1996.

\bibitem{p808}
{ITU-T Recommendation P.808}, ``Subjective evaluation of speech quality with a
  crowdsourcing approach,'' 2021.

\bibitem{pesq}
{ITU-T Recommendation P.862}, ``Perceptual evaluation of speech quality (pesq):
  An objective method for end-to-end speech quality assessment of narrow-band
  telephone networks and speech codecs,'' 2001.

\bibitem{polqa}
{ITU-T Recommendation P.863}, ``Perceptual objective listening quality
  prediction,'' 2018.

\bibitem{3quest}
{ETSI EG 202 393-3}, ``Speech processing, transmission and quality aspects
  (\uppercase{STQ}); speech quality performance in the presence of background
  noise, \uppercase{P}art 3: \uppercase{B}ackground noise transmission -
  \uppercase{O}bjective test methods,'' 2008.

\bibitem{nessler21_interspeech}
N.~Nessler, M.~Cernak, P.~Prandoni, and P.~Mainar, ``{Non-Intrusive Speech
  Quality Assessment with Transfer Learning and Subject-Specific Scaling},'' in
  \emph{Proc. Interspeech 2021}, 2021, pp. 2406--2410.

\bibitem{wawenets}
A.~A. Catellier and S.~D. Voran, ``Wawenets: A no-reference convolutional
  waveform-based approach to estimating narrowband and wideband speech
  quality,'' in \emph{ICASSP 2020 - 2020 IEEE International Conference on
  Acoustics, Speech and Signal Processing (ICASSP)}, 2020, pp. 331--335.

\bibitem{quality_net}
S.-W. Fu, Y.~Tsao, H.-T. Hwang, and H.-M. Wang, ``Quality-net: An end-to-end
  non-intrusive speech quality assessment model based on blstm,'' \emph{arXiv
  preprint arXiv:1808.05344}, 2018.

\bibitem{nisqa-v1}
G.~Mittag and S.~Möller, ``Non-intrusive speech quality assessment for
  super-wideband speech communication networks,'' in \emph{ICASSP 2019 - 2019
  IEEE International Conference on Acoustics, Speech and Signal Processing
  (ICASSP)}, 2019, pp. 7125--7129.

\bibitem{nisqa-v2}
\BIBentryALTinterwordspacing
G.~Mittag, B.~Naderi, A.~Chehadi, and S.~Möller, ``Nisqa: A deep
  cnn-self-attention model for multidimensional speech quality prediction with
  crowdsourced datasets,'' \emph{Interspeech 2021}, Aug 2021. [Online].
  Available: \url{http://dx.doi.org/10.21437/Interspeech.2021-299}
\BIBentrySTDinterwordspacing

\bibitem{dong_mos_multitask_attention}
X.~Dong and D.~S. Williamson, ``An attention enhanced multi-task model for
  objective speech assessment in real-world environments,'' in \emph{ICASSP
  2020 - 2020 IEEE International Conference on Acoustics, Speech and Signal
  Processing (ICASSP)}, 2020, pp. 911--915.

\bibitem{microsoft_nonintrusive}
A.~R. Avila, H.~Gamper, C.~Reddy, R.~Cutler, I.~Tashev, and J.~Gehrke,
  ``Non-intrusive speech quality assessment using neural networks,'' in
  \emph{ICASSP 2019-2019 IEEE International Conference on Acoustics, Speech and
  Signal Processing (ICASSP)}.\hskip 1em plus 0.5em minus 0.4em\relax IEEE,
  2019, pp. 631--635.

\bibitem{metricnet}
M.~Yu, C.~Zhang, Y.~Xu, S.-X. Zhang, and D.~Yu, ``{MetricNet: Towards Improved
  Modeling For Non-Intrusive Speech Quality Assessment},'' in \emph{Proc.
  Interspeech 2021}, 2021, pp. 2142--2146.

\bibitem{multitask-intelligibility-scene}
L.~Marcinek, M.~Stone, R.~Millman, and P.~Gaydecki,
  ``\BIBforeignlanguage{English}{N-mttl si model: Non-intrusive multi-task
  transfer learning-based speech intelligibility prediction model with scenery
  classification},'' in \emph{\BIBforeignlanguage{English}{Interspeech}}, Sep.
  2021.

\bibitem{room_acoustics_1}
P.~Callens and M.~Cernak, ``Joint blind room acoustic characterization from
  speech and music signals using convolutional recurrent neural networks,''
  \emph{arXiv preprint arXiv:2010.11167}, 2020.

\bibitem{room_acoustics_2}
P.~S. L{\'o}pez, P.~Callens, and M.~Cernak, ``A universal deep room acoustics
  estimator,'' in \emph{2021 IEEE Workshop on Applications of Signal Processing
  to Audio and Acoustics (WASPAA)}.\hskip 1em plus 0.5em minus 0.4em\relax
  IEEE, 2021, pp. 356--360.

\bibitem{vaswani}
\BIBentryALTinterwordspacing
A.~Vaswani, N.~Shazeer, N.~Parmar, J.~Uszkoreit, L.~Jones, A.~N. Gomez,
  L.~Kaiser, and I.~Polosukhin, ``Attention is all you need,'' \emph{CoRR},
  vol. abs/1706.03762, 2017. [Online]. Available:
  \url{http://arxiv.org/abs/1706.03762}
\BIBentrySTDinterwordspacing

\bibitem{caruana_multitask_learning}
\BIBentryALTinterwordspacing
R.~Caruana, ``Multitask learning,'' \emph{Machine Learning}, vol.~28, no.~1,
  pp. 41--75, Jul 1997. [Online]. Available:
  \url{https://doi.org/10.1023/A:1007379606734}
\BIBentrySTDinterwordspacing

\bibitem{kingma2014adam}
D.~P. Kingma and J.~Ba, ``Adam: A method for stochastic optimization,''
  \emph{arXiv preprint arXiv:1412.6980}, 2014.

\bibitem{challenge_evaluation_plan}
G.~Yi, W.~Xiao, Y.~Xiao, B.~Naderi, S.~Moller, G.~Mittag, R.~Cutler, Z.~Zhang,
  D.~S. Williamson, F.~Chen, F.~Yang, and S.~Shang, ``Conferencingspeech 2022
  challenge evaluation plan,'' 2022.

\bibitem{pstn}
G.~Mittag, R.~Cutler, Y.~Hosseinkashi, M.~Revow, S.~Srinivasan, N.~Chande, and
  R.~Aichner, ``{DNN No-Reference PSTN Speech Quality Prediction},'' in
  \emph{Proc. Interspeech 2020}, 2020.

\bibitem{p1401}
{ITU-T Recommendation P.1401}, ``Methods, metrics and procedures for
  statistical evaluation, qualification and comparison of objective quality
  prediction models,'' 2020.

\end{thebibliography}

\end{document}